\documentclass[preprint,authoryear,12pt]{elsarticle}

\usepackage{floatrow}
\usepackage{amsmath}
\usepackage{amssymb}

\journal{arXiv}

\usepackage{amsmath}
\usepackage{amsthm}

\newtheorem{theorem}{Theorem}
\newtheorem*{theorem*}{Theorem}
\newtheorem{lemma}{Lemma}
\newtheorem{remark}{Remark}

\usepackage{optidef}
\usepackage{subcaption}
\usepackage{nameref}
\usepackage{enumerate}
\usepackage{enumitem}

\allowdisplaybreaks

\usepackage{todonotes}

\usepackage{calc}
\usepackage{algorithm}
\usepackage{algpseudocode}
\usepackage{enumitem}
\usepackage{url}

\usepackage{hyperref}
\usepackage{breakurl}
\newcommand{\FWER}{\mbox{FWER}} 
\newcommand{\vt}{\vartheta}
\newcommand{\vp}{\varphi}
\newcommand{\aloc}{\alpha_{loc}}
\newcommand{\meff}{M_{\text{eff}}}
\newtheorem{defi}{Definition}
\newtheorem{modell}{Model}
\newtheorem{exam}{Example}
\newtheorem{algo}{Algorithm}
\newtheorem{scheme}{Scheme}

\DeclareMathOperator*{\argmaximum}{\arg\!\max}

\newcommand{\argmax}[1]{\ensuremath{\underset{#1}{\argmaximum}\ }}

\makeatletter

\def\namedlabel#1#2{\begingroup
    #2%
    \def\@currentlabel{#2}%
    \phantomsection\label{#1}\endgroup
}
\makeatother

\usepackage{mathtools}

\mathtoolsset{showonlyrefs}

\begin{document}

\begin{frontmatter}



\title{Optimizing effective numbers of tests by vine copula modeling}

\author{Nico Steffen}
\author{Thorsten Dickhaus\corref{cor1}}
\cortext[cor1]{Institute for Statistics, University of Bremen, 
P. O. Box 330 440, 28344 Bremen, Germany. Tel: +49 421 218-63651. E-mail address: dickhaus@uni-bremen.de (Thorsten Dickhaus).}

\address{Institute for Statistics, University of Bremen, Germany}

\begin{abstract}
In the multiple testing context, we utilize vine copulae for optimizing the effective number of tests. It is well known that for the calibration of multiple tests (for control of the family-wise error rate) the dependencies between the marginal tests are of utmost importance. It has been shown in previous work, that positive dependencies between the marginal tests can be exploited in order to derive a relaxed \v{S}id{\'{a}}k-type multiplicity correction. This correction can conveniently be expressed by calculating the corresponding "effective number of tests" for a given (global) significance level. This methodology can also be applied to blocks of test statistics so that the effective number of tests can be calculated by the sum of the effective numbers of tests for each block. In the present work, we demonstrate how the power of the multiple test can be optimized by taking blocks with high inner-block dependencies. The determination of those blocks will be performed by means of an estimated vine copula model. An algorithm is presented which uses the information of the estimated vine copula to make a data-driven choice of appropriate blocks in terms of (estimated) dependencies. Numerical experiments demonstrate the usefulness of the proposed approach.
\end{abstract}

\end{frontmatter}


\section{Introduction}\label{sec1}
Dependence modeling by means of copula functions has recently received a lot of attention in multiple testing, see \cite{stp-copulae}, \cite{Bodnar-Dickhaus-EJS},
\cite{Schmidt2014a}, \cite{Schmidt2014b}, \cite{UQ-FWER},
\cite{Cerqueti-Lupi2018}, \cite{Bernstein},
and Sections 2.2.4 and 4.4 of \cite{dick2014}. For example,  
\cite{stp-copulae} have explicitly shown that the copula approach
leads to the most general construction method for multivariate single-step multiple
tests under known univariate marginal null distributions of test
statistics or $p$-values, respectively. 

In the present work, we contribute to copula-based multiple testing by demonstrating how vine copula models (cf.\ \cite{czado-LNS2019} and references therein) can be used to optimize effective numbers of tests in the sense of \cite{dick2013} for control of the family-wise error rate (FWER). Assuming that the dependency structure among the test statistics is a nuisance parameter (of potentially infinite dimension), we propose to fit a regular vine copula model to the observed data. Under certain structural assumptions, this entails an approximation of the joint null distribution of the vector of test statistics, which can then be used to calibrate a multivariate multiple test for FWER control. By means of computer simulations, we will demonstrate that this strategy clearly improves existing approaches. In particular, choosing blocks of highly dependent test statistics by means of the estimated vine copula can lead to a substantial increase in statistical power when compared with a naively chosen block structure. 

The rest of the material is structured as follows. In Section \ref{sec2}, we introduce our basic statistical model, the concept of effective numbers of tests, and the vine copula modeling technique. Section \ref{sec3} contains our proposed methodology for combining these concepts. Some remarks on the implementation are provided in Section \ref{sec4}. Section \ref{sec5} presents numerical examples, and we conclude with a discussion in Section \ref{sec6}. 

\section{Notation and preliminaries}\label{sec2}

\subsection{Multiple testing}\label{sec21}
Throughout the work, we will assume the following statistical model.

\begin{modell}\label{grundmodell}
Let $n \in \mathbb{N}$ denote a sample size, and assume that we can observe stochastically independent and identically distributed (i.i.d.) random vectors $\mathbf{X}_1, \hdots,\mathbf{X}_n$, where $\mathbf{X}_k=(X_{k,1},...,X_{k,M})$ takes values in $\mathbb{R}^M$ for $1 \leq k \leq n$ and $M \in \mathbb{N}$. Altogether, this entails an observable random matrix 
\[
\mathbf{X}=\left(\begin{matrix}
\mathbf{X}_1\\
\vdots\\
\mathbf{X}_n
\end{matrix}\right)=\left(\begin{matrix}
X_{1,1} &\cdots & X_{1,M} \\

\vdots &   & \vdots \\
X_{n,1} &\cdots &X_{n,M}
\end{matrix}\right), 
\]
taking its values in the sample space $\mathcal{X}= \mathbb{R}^{n \times M}$. Assume that we have uncertainty about the distribution of $\mathbf{X}_1$. We express this by writing $\mathbf{X}_1 \sim P_{\boldsymbol\vartheta, C_\mathbf{X}}$, where $\boldsymbol\vartheta \in \Theta \subseteq \mathbb{R}^M$ is a parameter vector, such at each $\vartheta_j$ refers to the marginal distribution of $X_{1,j}$, $j \in I = \{1, \hdots, M\}$. Moreover, $C_\mathbf{X}$ denotes the copula of $\mathbf{X}_1$. For the distribution of the entire sample represented by $\mathbf{X}$, we write $\mathbb{P}_{\boldsymbol\vartheta, C_\mathbf{X}} = P_{\boldsymbol\vartheta, C_\mathbf{X}}^{\otimes n}$. Assume that we would like to test (simultaneously) $M$ null hypotheses $H_1, \hdots, H_M$, where each $H_j$ refers to $\vartheta_j$, $j \in I$. We may hence interpret each $H_j$ as a subset of $\mathbb{R}$. The corresponding alternative hypothesis will be denoted by $K_j = \mathbb{R} \setminus H_j$. For testing $H_j$ versus $K_j$, we assume that a real-valued test statistic $T_j: \mathbb{R}^n \to \mathbb{R}$ is at hand, where $T_j = T_j(X_{1, j}, \hdots, X_{n, j})$, $j \in I$. The vector of all $M$ test statistics will be denoted by $\boldsymbol{T} = (T_1, \hdots, T_M)^\top$. The multiple test based on $\boldsymbol{T}$ will be denoted by $\boldsymbol{\varphi} = (\varphi_1, \hdots, \varphi_M)^\top: \mathcal{X} \to \{0, 1\}^M$, where the event $\{\varphi_j = 1\}$ means that we reject $H_j$ in favor of $K_j$, $j \in I$. For the calibration of $\boldsymbol{\varphi}$, we aim at controlling the FWER, which is given by
\[
\FWER_{{\boldsymbol\vartheta, C_\mathbf{X}}}(\boldsymbol{\varphi}) = \bigcup_{j \in I_0} \{\varphi_j = 1\},
\] 
where $I_0 = I_0(\boldsymbol\vartheta) \subseteq I$ denotes the index set of true null hypotheses under 
$\boldsymbol\vartheta$. For a given constant $\alpha \in (0, 1)$, we say that $\boldsymbol{\varphi}$ controls the FWER at level $\alpha$ under $\boldsymbol\vartheta \in \Theta$ and  $C_\mathbf{X}$, if $\FWER_{{\boldsymbol\vartheta, C_\mathbf{X}}}(\boldsymbol{\varphi}) \leq \alpha$ holds true.
\end{modell}

Model \ref{grundmodell} is a standard multiple testing model in the context of studies with $M$ endpoints,  which are all measured for the same $n$ observational units; see, among many others, \cite{dick2013}, \cite{UQ-FWER}, and \cite{Bernstein}. Under Model \ref{grundmodell}, we make the following general assumptions.

\begin{description}
\item[\namedlabel{GA1}{(GA1)}] For all $j \in I$, the test statistic $T_j$ tends to larger values under the alternative $K_j$. We thus reject $H_j$ in favor of $K_j$ for large values of $T_j$.
\item[\namedlabel{GA2}{(GA2)}] The copula $C_\mathbf{X}$ is a nuisance parameter in the sense that it does not depend on $\boldsymbol{\vt}$.
\item[\namedlabel{GA3}{(GA3)}] There exists a parameter value $\boldsymbol{\vt}^*$ in the global null hypothesis $H_0 = \bigcap_{j=1}^M H_j$ which maximizes the FWER of the multiple test $\boldsymbol{\varphi}$ which is under consideration. Such a parameter value is often called a "least favorable (parameter) configuration", LFC for short.  
\item[\namedlabel{GA4}{(GA4)}] For all $j \in I$, the marginal distribution of $T_j$ under $\boldsymbol{\vt}^*$ is known, and it only depends on the $j$-th component $\vt_j^*$.
\end{description}

\begin{theorem}[Effective numbers of tests, Theorem 3.1 in \cite{dick2013}]\label{thm22}
Under our general assumptions \ref{GA1} - \ref{GA4}, let $C_\mathbf{X}$ be such, that $\boldsymbol{T}$ fulfills the MSM$_i$ property in the sense of Definition 2.2 of \cite{dick2013} for some $i \geq 1$ under $\boldsymbol\vt^*$.
Define critical values $\boldsymbol{c} = (c_1,\ldots,c_M)^\top \in \mathbb{R}^M$
such that  $\forall j\in I: \, \mathbb{P}_{\vt_j^{\ast}}(\vp_j=1) = \mathbb{P}_{\vt_j^{\ast}}(T_j>c_j)=\aloc$ for a fixed local significance level $\aloc\in(0,1)$ in each marginal.
Define also for $j \in I$:
\begin{eqnarray*}
\gamma_{j, 1} \equiv \gamma_{j, 1}(\boldsymbol{c}) &= &\mathbb{P}_{\vt_j^*}(T_j \leq c_j),\\
\gamma_{j, i} \equiv \gamma_{j,i}(\boldsymbol{c})&=&\mathbb{P}_{\boldsymbol\vartheta^*,C_\mathbf{X}} \left(T_j \leq c_j \mid \bigcap_{h=j-i+1}^{j-1} \{T_h \leq c_h \} \right), \text{~for~} 1 < i \leq j.
\end{eqnarray*} 
\begin{itemize}
\item [(i)]
In case of $i \leq 2$, set $\xi(i) = 0$. Otherwise, let 
$$
\xi(i) = \sum_{\ell=2}^{i-1} \frac{\log(\gamma_{\ell, \ell}(\boldsymbol{c}))}{\log(1-\aloc)}.
$$
Moreover, for every $i \leq j \leq M$, define
\[
\kappa^{(i)}_j \equiv \kappa^{(i)}_j(\aloc;T_1,\ldots,T_j) = \frac{\log(\gamma_{j, i}(\boldsymbol{c}))}{\log(1-\aloc)}. 
\]
 Then it holds
\begin{equation} \label{sidak-bound}
\FWER_{{\boldsymbol\vartheta^*,C_\mathbf{X}}}(\boldsymbol{\varphi})\leq 1-(1-\aloc)^{\meff^{(i)}}
\end{equation} 
for an "effective number of tests" of order $i$, given by
\[
\meff^{(i)}\equiv\meff^{(i)}(\aloc, \boldsymbol{T}) = 1 + \xi(i) + \sum_{j=i \vee 2}^M \kappa^{(i)}_j.
\]
\item[(ii)] Optimized bounds $\bar{\kappa}^{(i)}_j$ and $\bar{M}_\text{\tiny{eff.}}^{(i)}$:\\
If, for every permutation $\pi \in \mathcal{S}_M$, the MSM$_i$ property is preserved
if $\boldsymbol{T} = (T_1, \hdots, T_M)^\top$ is replaced by $(T_{\pi(1)}, \hdots, T_{\pi(M)})^\top$,
it is possible to optimize $\kappa^{(i)}_j$ and, consequently, $\meff^{(i)}$ in that the maximum
strength of positive dependence between $T_j$ and the preceding $T_h$, $1 \leq h \leq j-1$, is used.
For $i=2$, this leads to an optimized version
\[
\bar{\kappa}^{(2)}_j \equiv \bar{\kappa}^{(2)}_j(\aloc;T_1,\ldots,T_j) = \frac{\log(\max_{k<j}\mathbb{P}_{\boldsymbol\vartheta^*,C_\mathbf{X}}(T_j\leq c_j \, \vert \, T_k\leq c_k))}{\log(1-\aloc)}.
\]
An optimized effective number of tests of order $i$ is given by $\bar{M}_\text{\tiny{eff.}}^{(i)} = 1 + \xi(i) + \sum_{j=i \vee 2}^M \bar{\kappa}^{(i)}_j$.
\end{itemize}
\end{theorem}

\begin{remark}\label{rem-meff-practice} $ $
\begin{itemize}
\item[(a)] The MSM$_i$ property is a positive dependency property. In plain terms, it means that a particular test statistic $T_j$ tends to small values, given the information that $i-1$ test statistics $T_h$ with $h < j$ have realized small values.  
\item[(b)] The bound on the right-hand side of \eqref{sidak-bound} is of \v{S}id{\'{a}}k-type, where $M$ is replaced by $\meff^{(i)}$.
\item[(c)] It holds that $1 \leq \meff^{(i)} \leq M$. If $\meff^{(i)} < M$, this has the interpretation that we "effectively" only have to correct for $\meff^{(i)}$ tests, due to certain similarities between them.
\item[(d)] In practice, we have to find the value of $\aloc$ such that the right-hand side of \eqref{sidak-bound} equals the pre-defined global significance level $\alpha$. This can be achieved by starting with a reasonable upper bound for $\aloc$, and iteratively evaluating \eqref{sidak-bound} and decreasing $\aloc$ until the aforementioned equality holds (approximately).
\end{itemize}
\end{remark}

\subsection{Vine copulae}\label{sec22}
Here, we collect some essential definitions and properties of (regular) vines and vine copulae. For more details, see Chapter 5 in \cite{czado-LNS2019} and the references therein. 
\begin{defi}[Vine]\label{def_regvine}
Let $M \in \mathbb{N}$. The set $\mathcal{V} =\{\mathcal{T}_1, \hdots, \mathcal{T}_{M-1}\}$ is called a vine of  $M$ elements, where $E(\mathcal{V})=E_1 \cup \hdots \cup E_{M-1}$ denotes the set of edges of
$\mathcal{V}$, if
\begin{itemize}
\item[(i)] $\mathcal{T}_1$ is a connected tree with nodes $N_1=\{1,...,M\}$ and edges $E_1$, and
\item[(ii)] $\mathcal{T}_i$ is a tree with nodes $N_i=E_{i-1}$, for $2 \leq i \leq M-1$.
\end{itemize}
If it holds, in addition, that
\begin{enumerate}
\item[(iii)] $\#(a\bigtriangleup b)=2$ for all $2 \leq i \leq M-1$ and $\{a,b\}\in E_i$, where $\bigtriangleup$ denotes the symmetric difference,
\end{enumerate}
then $\mathcal{V}$ is called a regular vine (R-vine). As usual in graph theory, we call the number of edges which are connected to a particular node the degree of that node.
\end{defi}

\begin{defi}[Complete union, conditioning set, conditioned set] \label{def_vinenot}
Let $\mathcal{V}$ be a given vine of $M$ elements, and let $e_i \in E_i$ be a given edge. The set $U_{e_i}=\{n_1 \in N_1|\exists e_j \in E_j, j=1,...,i-1, \text{~such that~} n_1 \in e_1 \in \hdots \in e_{i-1}\in e_i \}\subseteq N_1$ is called the complete union of $e_i$. In words, $U_{e_i}$ denotes the set of nodes in the first tree $\mathcal{T}_1$ which can be "reached" from $e_i$. 
Letting $e_i=\{j,k\}$, we call \(D_{e_i}=U_j \cap U_k\) the conditioning set of $e_i$. 
Finally, the conditioned set $\mathcal{B}_{e_i}$ of $e_i$ is defined by
\[
\mathcal{B}_{e_i}= \mathcal{B}_{e_i,j}\cup \mathcal{B}_{e_i,k}=U_j \bigtriangleup U_k,
\]
where $\mathcal{B}_{e_i, \ell}=U_\ell \setminus D_{e_i}$ for $\ell \in \{j, k\}$.
\end{defi}

\begin{exam}\label{cd-vine-example}
Let $M = 4$. Figure \ref{fig-cd-vines} graphically displays two R-vine structures of four elements. 
\begin{figure}[htp]
\begin{center}
\includegraphics[width=0.6\textwidth]{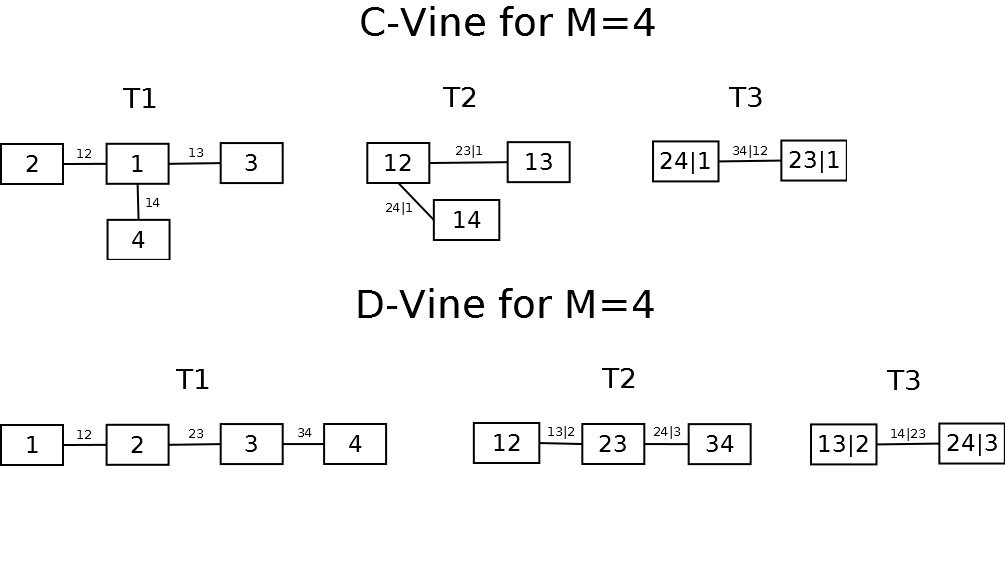}
\caption{$C$- and $D$-vines for $M = 4$.\label{fig-cd-vines}}
\end{center}
\end{figure}
\begin{itemize}
\item[(a)] The upper panel in Figure \ref{fig-cd-vines} displays a C-vine structure for $M = 4$. A C-vine is a regular vine fulfilling that every tree $\mathcal{T}_i$ has a node with degree $M-i$, for $1 \leq i \leq M-1$.
\item[(b)] The lower panel in Figure \ref{fig-cd-vines} displays a D-vine structure for $M = 4$. A D-vine is a regular vine fulfilling that all nodes in $\mathcal{T}_1$ have a degree of at most $2$.
\end{itemize}
Each edge in Figure \ref{fig-cd-vines} is labeled such, that the elements of the corresponding conditioning set are provided after the "$|$", and the elements of the corresponding conditioned set are provided before 
the "$|$". This kind of notation will be used throughout the remainder.
\end{exam}

\begin{defi}[(Regular) Vine distribution]\label{def_vinecop}
Let $\mathcal{V}=(\mathcal{T}_1, \hdots, \mathcal{T}_{M-1})$ be a given vine of $M$ elements. The vine distribution induced by $\mathcal{V}$ is given by the so-called "pair copula construction", meaning that a bivariate copula $C_e$ is attached to each edge $e \in E(\mathcal{V})=E_1 \cup \hdots \cup E_{M-1}$.
Formally, a triple $(\mathbf{F}, \mathcal{V},\mathbf{C}_2)$ is called a regular vine distribution, if 
$\mathbf{F}=(F_1,...,F_M)^\top$ is a vector of continuous and invertible cumulative distribution functions (cdfs) on $\mathbb{R}$, $\mathcal{V}$ is a regular vine in dimension $M$, and 
$\mathbf{C}_2= \{C_e: 1 \leq i \leq M-1, e \in E_i \}$ is a set of bivariate copula functions.

We say that the $\mathbb{R}^M$-valued random vector $(X_1,...,X_M)^\top$ possesses the regular vine distribution $(\mathbf{F}, \mathcal{V},\mathbf{C}_2)$, if $F_i$ is the marginal cdf of $X_i$ for all $1 \leq i \leq M$, and if $C_e$ is the (conditional) bivariate copula of $(X_{\mathcal{B}_{e,a}}, X_{\mathcal{B}_{e,b}})^\top$ given $\mathbf{X}_{D_e}$ for each edge $e = \{a,b\} \in E(\mathcal{V})$.

The tuple $(\mathcal{V},\mathbf{C}_2)$ will be referred to as a regular vine copula throughout the remainder.
\end{defi}

\begin{theorem}[Corollary 1 in \cite{buc2001}]\label{thm_buc}
Let $\mathcal{V}=(\mathcal{T}_1, \hdots, \mathcal{T}_{M-1})$ be a regular vine of $M$ elements. 
Assume that the (conditional) copula $C_{e_1,e_2|D_e}$  possesses a copula density $c_{e_1,e_2|D_e}$ for
each edge $e \in E(\mathcal{V})$ with conditioning set $D_e$ and conditioned elements $e_1$ and $e_2$. Furthermore, assume that the cdf $F_i$ admits a Lebesgue density $f_i$ for each $1 \leq i \leq M$. Then, there exists a unique probability distribution on $\mathbb{R}^M$ which has the Lebesgue density 
\begin{equation} \label{copuladichte}
f_{1 \hdots M}=\prod_{i=1}^M f_i \times \prod_{e\in E(\mathcal{V})} c_{e_1,e_2|D_e}(F_{e_1|D_e},F_{e_2|D_e}).
\end{equation}
Hence, under the aforementioned assumptions there exists an $\mathbb{R}^M$-valued random vector $(X_1,...,X_M)^\top$ possessing the regular vine distribution $(\mathbf{F}, \mathcal{V},\mathbf{C}_2)$, and $F_{e_\gamma|D_e}$ is the conditional cdf of $X_{e_\gamma}$ given $\mathbf{X}_{D_e}$, where $\gamma \in \{1, 2\}$
and $e \in E(\mathcal{V})$.
\end{theorem}

\section{Proposed methodology}\label{sec3}
The following lemma, the proof of which is deferred to \ref{sec7}, connects the concept of effective numbers of tests with copula theory.

\begin{lemma}\label{lemma31}
Let $j \in I$ and $1 < i \leq j$. For all $1 \leq h \leq j$, denote by $F_{T_h}$ the univariate marginal cdf of $T_h$ under $\boldsymbol\vt^*$, and assume that $F_{T_h}$ is continuous and strictly increasing on its support.
Furthermore, let $c_h = F_{T_h}^{-1}(1 - \alpha_{loc})$,  $1 \leq h \leq j$, for a fixed local significance level 
$\alpha_{loc} \in (0, 1)$. To avoid pathologies, assume that $\mathbb{P}_{\boldsymbol\vartheta^*,C_\mathbf{X}} \left(\bigcap_{h=j-i+1}^{j-1} \{T_h \leq c_h \} \right)>0$. Then we have, that
\begin{equation} \label{gamma-via-cop}
\gamma_{j,i}(\boldsymbol{c})=\frac{C_{T_{j-i+1}, \hdots, T_{j}}(1- \alpha_{loc}, \hdots,1- \alpha_{loc})}{C_{T_{j-i+1}, \hdots ,T_{j-1}}(1- \alpha_{loc}, \hdots,1- \alpha_{loc})}.
\end{equation}     
\end{lemma}

As mentioned in part (ii) of Theorem \ref{thm22}, it is advantageous for an optimization (in terms of statistical power) of the effective number of tests to find a structure / pattern in the degree of dependency among the test statistics $T_1, \hdots, T_M$. This means, that we aim at subdividing the total index set $I$ into blocks, such that the inner-block dependencies between test statistics are strong, while test statistics belonging to different blocks exhibit weak dependencies or are even stochastically independent. As argued in Section 5 of \cite{stange2016}, in some applications like, for instance, genetic association analyses, an appropriate block structure or at least appropriate block lengths can be deduced from expert knowledge about the experiment. An effective number of tests of appropriate order $i$ can then be calculated for every block 
$b$ separately. Letting $M_{\text{eff}, b}^{(i)}$ denote the effective number of tests of order $i$ in block 
$1 \leq b \leq B$, where $B$ is the total number of blocks, we can take the number
\begin{equation} \label{summe-meff}
M_{\text{eff}, 1}^{(i)} + M_{\text{eff}, 2}^{(i)} + \hdots + M_{\text{eff}, B}^{(i)}
\end{equation}
as a (conservative) approximation of the optimized total effective number of tests $\bar{M}_\text{\tiny{eff.}}^{(i)}$ of order $i$. In this, the term "conservative" means, that under MSM$_i$ the sum in \eqref{summe-meff} is guaranteed to be not smaller than $\bar{M}_\text{\tiny{eff.}}^{(i)}$; see Section 5 in \cite{stange2016} for further details.

Our proposed methodology is to apply vine copula modeling for (i) finding the appropriate block structure, and (ii) estimating the copulae appearing in \eqref{gamma-via-cop}. The underlying pair copula construction can be exploited in a greedy-style algorithmic manner to determine an appropriate block structure. This property makes the vine approach particularly well-suited in our context. In the remainder of this section, we explain the two steps indicated above in more detail.

For automated model selection and pair copula estimation, we employ Algorithm 3.1 of \cite{diss2013}, referred to as "The Di\ss{}mann Algorithm" in Section 8.3 of \cite{czado-LNS2019}. This algorithm delivers an estimate 
$\hat{\mathcal{V}}$ of the vine structure $\mathcal{V}$ underlying the data, as well as an estimate $\hat{\mathbf{C}}_2$ of the family $\mathbf{C}_2$ of pair copulae underlying the construction in \eqref{copuladichte}. Furthermore, for regularization purposes, we choose a truncation level $K \leq M$ and finally work with the approximation 
\begin{equation} \label{truncated-vine}
\hat{f}_{1 \hdots M}=\prod_{i=1}^M f_i \times \prod_{i=1}^K\prod_{e\in \hat{E}_i} \hat{c}_{e_j,e_k|D_e}(\hat{F}_{e_j|D_e}, \hat{F}_{e_k|D_e}),
\end{equation}
where the $\hat{E}_i$'s refer to the estimated vine structure $\hat{\mathcal{V}}$. A formal, information criterion-based method for choosing $K$ has been proposed by \cite{Nagler2019}. However, from our experience, the choice $K \equiv 2$ often works well in practice.

\begin{remark} $ $
\begin{itemize}
\item[(i)] Notice, that we do not have to estimate the (univariate) marginal densities $f_i$ for $1 \leq i \leq M$ in \eqref{truncated-vine}, because we have to calibrate the multiple test under the LFC $\boldsymbol{\vt}^* \in H_0$, and under $\boldsymbol{\vt}^*$ the marginal distributions of the test statistics are assumed to be known, see \ref{GA4}.
\item[(ii)] Comparing \eqref{truncated-vine} with \eqref{copuladichte} and noticing that the density of the independence (or: product) copula is identically equal to one on the unit cube, it becomes clear that in \eqref{truncated-vine} only the first $K$ (estimated) trees of $\hat{\mathcal{V}}$ are explicitly taken into account. In the remaining $M-1-K$ trees, all pair copulae are set to the independence copula. This strategy is justified, because the Di\ss{}mann Algorithm is designed to capture the most pronounced dependencies in the first trees.
\item[(iii)] The estimated (joint) density $\hat{f}_{1 \hdots M}$ refers to the distribution of $\mathbf{X}_1$. For calibrating the multiple test $\boldsymbol{\varphi}$, though, we need the null distribution of the vector $\boldsymbol{T}$ of test statistics. However, since $\boldsymbol{T} = \boldsymbol{T}(\mathbf{X}_1, \hdots, \mathbf{X}_n)$ is a given function of the (i.i.d.) data vectors $\mathbf{X}_1, \hdots, \mathbf{X}_n$, the dependency structure among the components $X_{1,1},...,X_{1,M}$ of $\mathbf{X}_1$ already determines the dependency structure among $T_1, \hdots, T_M$. Even if the mapping $\boldsymbol{T}$ is complicated, we can approximate the joint distribution of the random vector $\boldsymbol{T}(\mathbf{X}_1, \hdots, \mathbf{X}_n)$ under $\boldsymbol{\vt}^* \in H_0$ with arbitrary precision by means of a Monte Carlo simulation, once the dependency structure among the components of $\mathbf{X}_1$ has been estimated. 
\end{itemize}
\end{remark}

Based on the estimated quantities $\hat{\mathcal{V}}$ and $\hat{\mathbf{C}}_2$, we propose the following algorithm for finding appropriate blocks for the operationalization of \eqref{summe-meff}. In Algorithm \ref{algo-groups}, we assume that the blocks (or groups) are all of (approximately) equal size.

\begin{algo}[Greedy algorithm for determining a grouping of the $M$ test statistics]\label{algo-groups} $ $
\begin{itemize}
\item[$\boldsymbol{Input}$] The estimated quantities $\hat{\mathcal{V}}$ and $\hat{\mathbf{C}}_2$, and the targeted group size.
\item[$\boldsymbol{Output}$] Grouping of the $M$ test statistics, which correspond to the nodes in the first tree in $\hat{\mathcal{V}}$.
\end{itemize}

\begin{itemize}
\item[1)] Find the pair of coordinates with largest estimated Kendall's $\tau$ coefficient 
(according to $\hat{\mathbf{C}}_2$), and assign the two corresponding nodes to Group 1.
\item[2)] Find all nodes, which share an edge (according to $\hat{\mathcal{V}}$) with a node in Group 1, but are themselves not in Group 1 (yet). We call these nodes the neighbors.
\item[3)] Choose the neighbor with the strongest dependency with Group 1, and assign this neighbor to Group 1. 
This means, that we find
\begin{equation} \label{maxnachbar}
\argmax{n \in \text{Set of neighbors}} |\hat{\tau}_{n,g(n)}|+\sum_{i \in H(g(n))}|\hat{\tau}_{n,i|g(n)}|,
\end{equation}
where $g(n)$ denotes the neighboring node of $n$ from Group 1, and $H(g(n))$ denotes the set of nodes from Group 1 which are neighbors of $g(n)$. In \eqref{maxnachbar}, $\hat{\tau}_{i, j}$ denotes the estimated (unconditional) Kendall's $\tau$ coefficient of $X_{1, i}$ and $X_{1, j}$, and $\hat{\tau}_{j,i|k}$ denotes the estimated conditional Kendall's $\tau$ coefficient of $X_{1, j}$ and $X_{1, i}$ given $X_{1, k}$.
\item[4)] Repeat 3), until Group 1 has reached the targeted group size.
\item[5)] For Group 2 until Group $B$ (last group), carry out steps 1) to 4) analogously by considering only those nodes, which have not been assigned to any group yet. If no neighbors are found, go to the next group.
\item[6)] If there are still nodes left which have not been assigned to any group yet, assign them randomly to those groups which have not yet reached the targeted group size.
\end{itemize}
\end{algo}

\begin{remark}\label{remark-neighbors} $ $
\begin{itemize}
\item[(i)] The neighboring node $g(n)$ appearing in \eqref{maxnachbar} is uniquely determined, because the tree contains no cycle. 
\item[(ii)] When constructing Group 1, it is guaranteed that a neighbor can be found.
\item[(iii)] In the case that $B$ is fixed in advance, the targeted group size is $\lfloor M / B\rfloor$.
\item[(iv)] For Algorithm \ref{algo-groups}, only the first two trees in $\hat{\mathcal{V}}$, together with their corresponding pair copulae from $\hat{\mathbf{C}}_2$, are needed.
\item[(v)] For any given copula $C$ on $[0, 1]^2$, the corresponding Kendall's $\tau$ coefficient is given by
\[
\tau(C) = 4\int\int_{[0,1]^{2}}C(u,v)dC(u,v)-1=4\mathbb{E}[C(U,V)]-1,
\]
with $(U,V)^{\top}\sim C$.
\end{itemize}
\end{remark}

\section{Details on the implementation}\label{sec4}
Summarizing the proposed methodology presented in Sections \ref{sec2} and \ref{sec3}, we obtain the following data analysis workflow.

\begin{scheme}\label{scheme1}
Given are the realized data matrix $\mathbf{X} = \mathbf{x}$, the null hypotheses $H_1, \hdots, H_M$, the mappings (test statistics) $T_1, \hdots, T_M$, their univariate marginal cdfs $(F_{T_j}: 1 \leq j \leq M)$ under $\boldsymbol \vt^*$, the FWER level $\alpha$, the order $i$ for the effective numbers of tests, and the number $B$ of blocks. 
\begin{itemize}
\item[1)] In order to have (approximately) marginally uniformly distributed data as input for the Di\ss{}mann Algorithm, we transform the data points $(x_{k, j})_{\substack{1\leq k \leq n\\1 \leq j \leq M}}$ with their empirical marginal cdfs, meaning that we set
\[
u_{k,j}=\hat{F}_{j}(x_{k,j}), \; 1 \leq k \leq n, 1 \leq j \leq M.
\]
\item[2)] We apply the  Di\ss{}mann Algorithm to $(u_{k,j}: 1 \leq k \leq n, 1 \leq j \leq M)$ obtained in Step 1), and receive the estimated quantities $\hat{\mathcal{V}}$ and $\hat{\mathbf{C}}_2$. The Di\ss{}mann Algorithm also computes and outputs all estimated (conditional and unconditional) Kendall's $\tau$ coefficients pertaining to $\hat{\mathbf{C}}_2$; cf.\ Part (v) of Remark \ref{remark-neighbors}.
\item[3)] We apply Algorithm \ref{algo-groups} to the estimated quantities obtained in Step 2), and thereby determine the $B$ blocks for operationalizing \eqref{summe-meff}.
\item[4)] We carry out a Monte Carlo simulation for approximating the joint distribution of $T_1, \hdots, T_M$under the global null hypothesis $H_0$. In this simulation, we combine the univariate marginal cdfs $(F_{T_j}: 1 \leq j \leq M)$ under $\boldsymbol{\vartheta}^*$ with  the estimated vine copula from Step 2).
\item[5)] We compute the block-wise effective numbers of tests $M_{\text{eff}, 1}^{(i)}, \hdots, M_{\text{eff}, B}^{(i)}$ of order $i$ as well as the critical values $c_1,...,c_M$ based on the estimated joint null distribution from Step 4); see Theorem \ref{thm22} and Remark \ref{rem-meff-practice}.
\item[6)] We reject the global null hypothesis $H_0$, iff there exists an $1 \leq j \leq M$ with $T_j(x_{1, j}, \hdots, x_{n, j}) >c_j$. Furthermore, we reject all individual null hypotheses $H_j$ with $T_j(x_{1, j}, \hdots, x_{n, j}) >c_j$, $1 \leq j \leq M$.
\end{itemize}
\end{scheme}

In the remainder of this section, we briefly describe how we have implemented this workflow in the statistical computing environment \verb=R= (\url{https://www.r-project.org/}).
\subsection{Implementation of the Di\ss{}mann Algorithm} \label{sec41}
The  Di\ss{}mann Algorithm is included in the \verb=R= package \verb=VineCopula= (\url{https://cran.r-project.org/web/packages/VineCopula/}); see the function \linebreak[4] \texttt{RVineStructureSelect} in that package.  As one argument, the transformed data $(u_{k,j}: 1 \leq k \leq n, 1 \leq j \leq M)$ are required. As another argument, copula families are required, from which the bivariate copulae appearing in  $\hat{\mathbf{C}}_2$ are chosen. For our numerical experiments described in Section \ref{sec5}, we have taken the following families.
\begin{itemize}
\item[0)] Independence copula (product copula),
\item[1)] Gaussian copula family,
\item[3)] Clayton copula family,
\item[4)] Gumbel copula family,
\item[5)] Frank copula family,
\item[6)] Joe copula family,
\end{itemize} 
together with their rotated versions. The numbering in the above list corresponds to that in the \verb=R= package \verb=VineCopula=. The package also offers further families, but we have worked only with the aforementioned ones. Finally, the function \verb=RVineStructureSelect= requires the specification of the truncation level $K$. In our experiments, we have set $K=2$; cf.\ Part (iv) of Remark \ref{remark-neighbors}.
\subsection{Implementation of the other steps in Scheme \ref{scheme1}} \label{sec42}
A custom implementation of Algorithm \ref{algo-groups} is available from the authors upon request. Notice, that all required quantities for Algorithm \ref{algo-groups} (namely, $\hat{\mathcal{V}}$, $\hat{\mathbf{C}}_2$, as well as the estimated (conditional and unconditional) Kendall's $\tau$ coefficients pertaining to $\hat{\mathbf{C}}_2$) are delivered by the Di\ss{}mann Algorithm. Hence, it essentially remains to code the neighbor search and the evaluation of \eqref{maxnachbar}.

For the simulation in Step 4) of Scheme \ref{scheme1}, we have used the function \verb=RVineSim= from the 
\verb=R= package \verb=VineCopula=. This function generates pseudo-random vectors from the estimated vine copula. Combining this with the principle of quantile transformation yields pseudo-random vectors which behave like realizations of $\boldsymbol{T}$ under $\boldsymbol{\vartheta}^*$.

Many further resources for working with vine copulae can be found at \url{http://www.vine-copula.org/}.

\section{Numerical experiments}\label{sec5}
\subsection{Multivariate Gaussian model}\label{sec51}
In our first numerical example, we let $M=15$. We assume that $\mathbf{X}_1$ follows the $15$-variate normal distribution with mean vector $\boldsymbol{\vartheta}$ and covariance matrix $\Sigma$. In our simulations, 
we set $\vartheta_j = 0$ for $1 \leq j \leq 11$ and $\vartheta_j =0.15$ for $12 \leq j \leq 15$. The covariance matrix is given by
\setcounter{MaxMatrixCols}{20}
\begin{equation*}
\Sigma=\begin{pmatrix}
 1 & 0 & 0 & \frac{9}{10} & 0 & 0 & \frac{9}{10} & 0 & 0 & \frac{9}{10} & 0 & 0 & \frac{9}{10} & 0 & 0\\
 0 & 1 & 0 & 0 & \frac{9}{10} & 0 & 0 & \frac{9}{10} & 0 & 0 & \frac{9}{10} & 0 & 0 & \frac{9}{10} & 0\\
0 & 0 & 1 & 0 & 0 & \frac{9}{10} & 0 & 0 & \frac{9}{10} & 0 & 0 & \frac{9}{10} & 0 & 0 & \frac{9}{10}\\
\frac{9}{10} & 0 & 0 & 1 & 0 & 0 & \frac{9}{10} & 0 & 0 & \frac{9}{10} & 0 & 0 & \frac{9}{10} & 0 & 0\\
0 & \frac{9}{10} & 0 & 0 & 1 & 0 & 0 & \frac{9}{10} & 0 & 0 & \frac{9}{10} & 0 & 0 & \frac{9}{10} & 0\\
0 & 0 & \frac{9}{10} & 0 & 0 & 1 & 0 & 0 & \frac{9}{10} & 0 & 0 & \frac{9}{10} & 0 & 0 & \frac{9}{10}\\
\frac{9}{10} & 0 & 0 & \frac{9}{10} & 0 & 0 & 1 & 0 & 0 & \frac{9}{10} & 0 & 0 & \frac{9}{10} & 0 & 0\\
0 & \frac{9}{10} & 0 & 0 & \frac{9}{10} & 0 & 0 & 1 & 0 & 0 & \frac{9}{10} & 0 & 0 & \frac{9}{10} & 0\\
0 & 0 & \frac{9}{10} & 0 & 0 & \frac{9}{10} & 0 & 0 & 1 & 0 & 0 & \frac{9}{10} & 0 & 0 & \frac{9}{10}\\
\frac{9}{10} & 0 & 0 & \frac{9}{10} & 0 & 0 & \frac{9}{10} & 0 & 0 & 1 & 0 & 0 & \frac{9}{10} & 0 & 0\\
0 & \frac{9}{10} & 0 & 0 & \frac{9}{10} & 0 & 0 & \frac{9}{10} & 0 & 0 & 1 & 0 & 0 & \frac{9}{10} & 0\\
0 & 0 & \frac{9}{10} & 0 & 0 & \frac{9}{10} & 0 & 0 & \frac{9}{10} & 0 & 0 & 1 & 0 & 0 & \frac{9}{10}\\
\frac{9}{10} & 0 & 0 & \frac{9}{10} & 0 & 0 & \frac{9}{10} & 0 & 0 & \frac{9}{10} & 0 & 0 & 1 & 0 & 0\\
0 & \frac{9}{10} & 0 & 0 & \frac{9}{10} & 0 & 0 & \frac{9}{10} & 0 & 0 & \frac{9}{10} & 0 & 0 & 1 & 0\\
0 & 0 & \frac{9}{10} & 0 & 0 & \frac{9}{10} & 0 & 0 & \frac{9}{10} & 0 & 0 & \frac{9}{10} & 0 & 0 & 1
\end{pmatrix}.
\end{equation*}
The marginal test problems of interest are assumed to be $H_j: \,\{ \vartheta_j = 0 \}$ versus $K_j: \,\{ \vartheta_j \neq 0 \}$ for $1 \leq j \leq 15$. The vector $\boldsymbol{T} = (T_1, \hdots, T_{15})^\top$ is given by $T_j= |\sqrt{n}\bar{X}_{n,j}|$, where $\bar{X}_{n,j}= n^{-1} \sum_{k=1}^n X_{k,j}$, for $1 \leq j \leq 15$. We let $\alpha = 5\%$. Making use of Propositions 4.1 and 4.2 in \cite{dick2014}, it can be shown that $\boldsymbol{T}$ fulfills the MSM$_2$ property under $\boldsymbol\vt^* = \boldsymbol{0} \in \mathbb{R}^{15}$, and that this property is preserved under coordinate permutations.

Analyzing the structure of $\Sigma$, we see that there are three blocks of highly correlated coordinates, namely $(1,4,7,10,13)$, $(2,5,8,11,14)$, and $(3,6,9,12,15)$. The goals of our computer simulations are to assess (i) how reliably our proposed methodology can identify these blocks, and (ii) how much gain in statistical power can be achieved by exploiting the dependency structure. We performed $400$ simulation runs for sample sizes $n \in \{100,200,300\}$. The number of groups has been set to $B = 3$, with a targeted group size of five per group.

\begin{figure}[htp]
\begin{center}
\includegraphics[width=\textwidth]{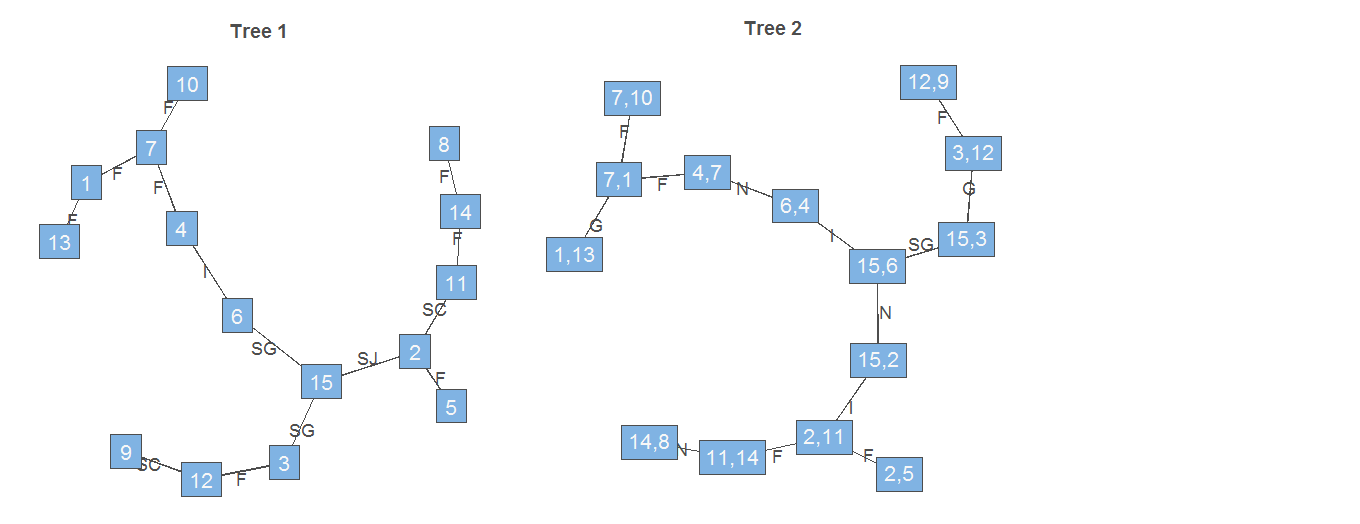}
\end{center}
\caption{The two estimated trees in $\hat{\mathcal{V}}$ for one simulation run with $n=300$ under the model from Section \ref{sec51}. The graphs have been produced by the function \texttt{RVineTreePlot} from the \texttt{R} package \texttt{VineCopula}.\label{fig-gausstrees}}
\end{figure}

\begin{figure}[htp]
\begin{center}
\includegraphics[width=\textwidth]{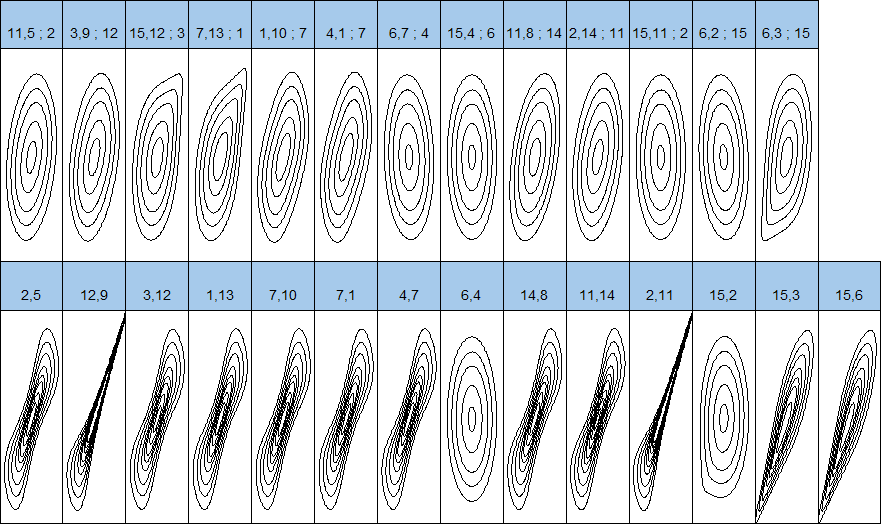}
\end{center}
\caption{Contour lines of the estimated pair copulas in $\hat{\mathbf{C}}_2$ for one simulation run with $n=300$ under the model from Section \ref{sec51}. The graphs have been produced by the function \texttt{contour} from the \texttt{R} package \texttt{VineCopula}. \label{fig-gausscontours}}
\end{figure}

For one particular simulation run with $n=300$, Figure \ref{fig-gausstrees} displays the two estimated trees in $\hat{\mathcal{V}}$. Furthermore, Figure \ref{fig-gausscontours} displays the contour lines of the estimated pair copulas in $\hat{\mathbf{C}}_2$.

Tables \ref{Table-Gauss1} - \ref{Table-Gauss3} display our (averaged) simulation results. Since all test statistics have the same null distribution, we have chosen $c_1 = c_2 = \hdots = c_{15} = c$. The rows labeled 
"fixed groups" refer to the fixed group structure $(1,2,3,4,5)$, $(6,7,8,9,10)$, and $(11,12,13,14,15)$.

\begin{table}[htp]
\centering
  \begin{tabular}{ l  c  c  c } 
      Sample size & 100 & 200 & 300  \\ 
    \v{S}id{\'{a}}k correction &15 &15 &15\\ 
		fixed groups  & 11.84 & 11.79 & 11.72 \\ 
     chosen groups & 8.63 & 8.75 & 8.8 \\ 
  \end{tabular}
	\caption{Average values of  $M_{\text{eff}, 1}^{(2)} + M_{\text{eff}, 2}^{(2)} + M_{\text{eff}, 3}^{(2)}$ in a computer simulation with $400$ simulation runs under the model from Section \ref{sec51}.}\label{Table-Gauss1}
\end{table}
\begin{table}[htp]
\centering
  \begin{tabular}{ l  c  c  c }
     Sample size &100 &200 &300  \\ 
		 \v{S}id{\'{a}}k correction &2.928 &2.928 & 2.928\\	
     fixed groups & 2.857 & 2.856 & 2.854 \\ 
     chosen groups & 2.755 & 2.759 & 2.761 \\  
  \end{tabular}
	\caption{Average values of $c$ in a computer simulation with $400$ simulation runs under the model from Section \ref{sec51}. The corresponding local significance level can be computed as $\aloc = 2 (1 - \Phi(c))$, where $\Phi$ denotes the cdf of the standard normal distribution on $\mathbb{R}$.}\label{Table-Gauss2}
\end{table}
\begin{table}[htp]
\centering
  \begin{tabular}{ l  c  c  c }
     Sample size &100 &200 &300 \\
		 \v{S}id{\'{a}}k correction &9 &22.625 & 37.875\\
     fixed groups & 9.75 & 24.75 & 40.6875 \\ 
     chosen groups & 11.4375 & 27.125 & 44.25 \\  
  \end{tabular}
	\caption{Empirical powers in per cent in a computer simulation with $400$ simulation runs under the model from Section \ref{sec51}.}\label{Table-Gauss3}
\end{table}

The results in Tables \ref{Table-Gauss1} - \ref{Table-Gauss3} clearly demonstrate the advantage of choosing the blocks in a data-driven manner. Our proposed methodology leads to a decrease in the effective numbers of tests and in turn to an increase in statistical power, when compared with the setup with fixed groups. 

Finally, we have also simulated under $\boldsymbol\vt^* = \boldsymbol{0} \in \mathbb{R}^{15}$, in order to assess how well the FWER level $\alpha = 5\%$ is kept when applying our proposed methodology. As displayed in Table \ref{Table-Gauss4}, we have found no indication for a violation of the FWER level.

\begin{table}[htp]
\centering
  \begin{tabular}{ l  c  c  c } 
      Sample size &100 &200 &300 \\
		 \v{S}id{\'{a}}k correction &1.25 &2.75 & 2.5\\	
     fixed groups & 3.25 & 3 & 3.25 \\ 
     chosen groups & 4 & 3.75 & 4 \\  
  \end{tabular}
	\caption{Empirical FWERs in per cent in a computer simulation with $400$ simulation runs under the model from Section \ref{sec51}. The simulation has been performed under $\boldsymbol\vt^* = \boldsymbol{0} \in \mathbb{R}^{15}$.}\label{Table-Gauss4}
\end{table}

\subsection{Vine copula model}\label{sec52}
\begin{figure}[htp]
\begin{center}
\includegraphics[width=\textwidth]{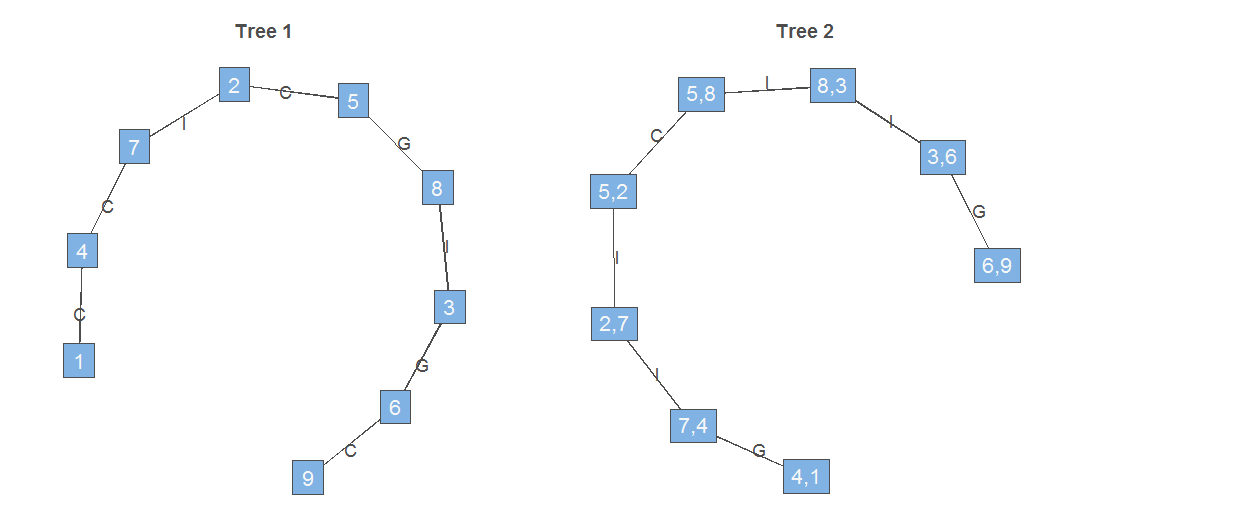}
\end{center}
\caption{The first two trees in the D-Vine $\mathcal{V}$ which has been used in the simulation study of Section \ref{sec52}.\label{fig-vine-Sec52}}
\end{figure}

In our second numerical example, we let $M=9$, and we assume that the Lebesgue density of $\mathbf{X}_1$ on $\mathbb{R}^9$ is given by
\begin{equation}\label{DGP-Sec52}
f_{\mathbf{X}_1} = \prod_{i=1}^9 f_i \times \prod_{e\in E(\mathcal{V})} c_{e_1,e_2|D_e}(F_{e_1|D_e},F_{e_2|D_e});
\end{equation}
cf.\ \eqref{copuladichte}. The marginal densities $f_1,\hdots,f_9$ are Lebesgue densities of univariate normal distributions with unit variance. The expected values of these normal distributions are set to zero in the first five coordinates and the remaining ones are set to $0.15$.  The vine $\mathcal{V}$ utilized in \eqref{DGP-Sec52} is a D-Vine with truncation level $K=2$, and its structure is displayed in Figure \ref{fig-vine-Sec52}.

In Table \ref{table-copulae-Sec52}, we list the copula families utilized in $\mathcal{V}$, together with the values of the associated copula parameters.
$ $\\

\begin{figure}[htp]
\begin{center}
\includegraphics[width=0.8\textwidth]{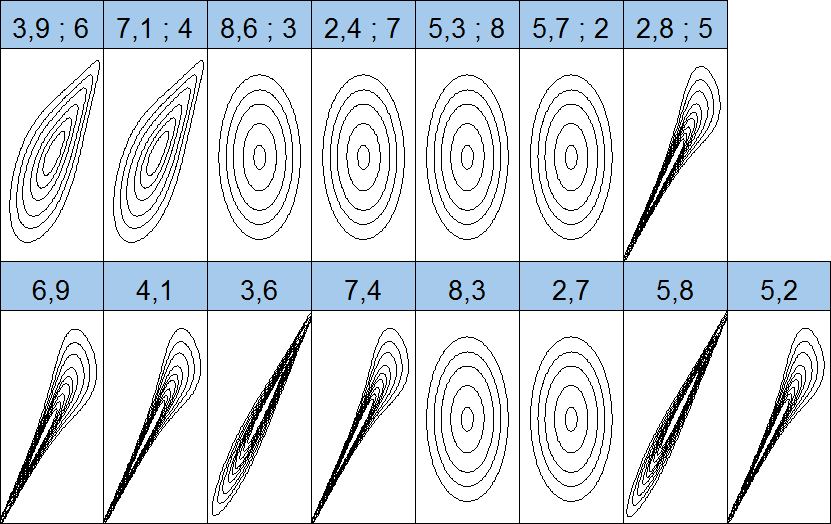}
\end{center}
\caption{Contour plots of the pair copulae which have been used in the simulation study of Section \ref{sec52}.\label{fig-contour-Sec52}}
\end{figure}

\begin{table}[htp]
\centering
\begin{tabular}{c c c c }
Tree & Nodes & Copula family & Copula parameter\\
  \hline			
  1 & 1,4 & Clayton & 11 \\
   & 4,7 & Clayton & 12\\
   & 2,5 & Clayton & 12 \\
   & 5,8 & Gumbel & 8\\
   & 3,6 & Gumbel & 7\\
   & 6,9 & Clayton & 8\\
  \hline  
  2 & 1,7|4 &  Gumbel & 2\\
    & 2,8|5 &  Clayton & 11\\
    & 3,9|6 &  Gumbel & 2\\
  \hline
\end{tabular}
\caption{Copula families and copula parameters for the simulation study of Section \ref{sec52}.}\label{table-copulae-Sec52}
\end{table}
To all nodes which do not appear in Table \ref{table-copulae-Sec52}, the independence copula has been assigned. Hence, we have the three independent blocks $(1,4,7)$, $(2,5,8)$, and $(3,6,9)$ of three coordinates each in the data-generating process for the distribution of $\mathbf{X}_1$. For a further illustration, Figure \ref{fig-contour-Sec52} displays the corresponding contour plots.

We assume, that $\vartheta_j$ is given as the marginal expected value in coordinate $1 \leq j \leq 9$, and that the test problem of interest is given by $H_j: \{ \vartheta_j \leq 0 \}$ versus $K_j: \{ \vartheta_j > 0 \}$ for $1 \leq j \leq M=9$. The test statistics are given by $T_j= \sqrt{n}\bar{X}_{n,j}$, where $\bar{X}_{n, j}= {n}^{-1} \sum_{k=1}^n X_{k,j}$ for $1 \leq j \leq 9$. Hence, $\boldsymbol\vt^* = \boldsymbol{0} \in \mathbb{R}^{9}$. Under $\boldsymbol\vt^*$, each $T_j$ marginally possesses the standard normal distribution on $\mathbb{R}$ (leading to choosing the same critical value $c$ for each $T_j$), while $T_j$ has a shifted normal distribution under the alternative $K_j$. In terms of the marginal tests, the only difference to the setup in Section \ref{sec51} is, that we now carry out one-sided $Z$-tests instead on two-sided ones. Furthermore, the dependency structure of $\mathbf{X}_1$ is now much more involved, such that a simple check of the validity of the MSM$_2$ property is not straightforward here. However, notice that the copula families appearing in Table \ref{table-copulae-Sec52} are only capable of expressing positive dependencies, indicating that MSM$_2$ is likely to be fulfilled in this simulation. In particular, the covariance matrix of $\mathbf{X}_1$ has only 
non-negative entries.

In analogy to Section \ref{sec51}, we display our (averaged) simulation results in Tables \ref{Table-Vine1} - \ref{Table-Vine4}.

\begin{table}[htp]
\centering
  \begin{tabular}{ l  c  c  c } 
      Sample size & 100 & 200 & 300  \\ 
    \v{S}id{\'{a}}k correction &9 &9 &9\\ 
		fixed groups  & 7.68 & 7.78 & 7.78\\ 
     chosen groups & 4.95 & 4.91 & 4.88\\ 
  \end{tabular}
	\caption{Average values of  $M_{\text{eff}, 1}^{(2)} + M_{\text{eff}, 2}^{(2)} + M_{\text{eff}, 3}^{(2)}$ in a computer simulation with $400$ simulation runs under the model from Section \ref{sec52}.}\label{Table-Vine1}
\end{table}
\begin{table}[htp]
\centering
  \begin{tabular}{ l  c  c  c }
     Sample size &100 &200 &300  \\ 
		 \v{S}id{\'{a}}k correction &2.531 &2.531 & 2.531\\	
     fixed groups & 2.478 & 2.482 & 2.482\\ 
     chosen groups & 2.318 & 2.315 & 2.313 \\  
  \end{tabular}
	\caption{Average values of $c$ in a computer simulation with $400$ simulation runs under the model from Section \ref{sec52}. The corresponding local significance level can be computed as $\aloc = 1 - \Phi(c)$, where $\Phi$ denotes the cdf of the standard normal distribution on $\mathbb{R}$.}\label{Table-Vine2}
\end{table}
\begin{table}[htp]
\centering
  \begin{tabular}{ l  c  c  c }
     Sample size &100 &200 &300 \\
		 \v{S}id{\'{a}}k correction & 15.625 & 32.625 & 52.3125\\
     fixed groups               & 16.75 & 34.25 & 54.8125\\ 
     chosen groups              & 21.875 & 40.375 & 60.8125 \\  
  \end{tabular}
	\caption{Empirical powers in per cent in a computer simulation with $400$ simulation runs under the model from Section \ref{sec52}.}\label{Table-Vine3}
\end{table}

\begin{table}[htp]
\centering
  \begin{tabular}{ l  c  c  c } 
      Sample size &100 &200 &300 \\
		 \v{S}id{\'{a}}k correction  &3.5 &1.25 &2.25\\	
     fixed groups & 3.5 & 1.25 & 3.25  \\ 
     chosen groups & 4.75 & 2.75 & 4.25\\  
  \end{tabular}
	\caption{Empirical FWERs in per cent in a computer simulation with $400$ simulation runs under the model from Section \ref{sec52}. The simulation has been performed under $\boldsymbol\vt^* = \boldsymbol{0} \in \mathbb{R}^{9}$.}\label{Table-Vine4}
\end{table}

\section{Conclusion}\label{sec6}
We have proposed a vine copula-based construction method for multivariate multiple tests. The main advantage of the vine copula estimation approach is, that the tree structure in $\hat{\mathcal{V}}$ straightforwardly allows for choosing appropriate blocks for a block-wise evaluation of the effective numbers of tests. In the computer simulations presented in Section \ref{sec5}, the dependency structure was explicitly given. Notice, however, that the workflow from Scheme \ref{scheme1} is data-driven in the sense, that the pair copulae are chosen from a large pool of copula families on the basis of the sample only, without relying on any prior information about the type of dependencies among the test statistics. This is particularly useful in cases with a moderate or high dimensionality $M$, when it is typically infeasible to model (pair) copulae of the data explicitly. Due to the combinatorial explosion involved in the vine model selection, we consider 
$M = 15$ (cf.\ Section \ref{sec51}) or $M = 9$ (cf.\ Section \ref{sec52}) already as quite high dimensionalities in our context.

There are several possible extensions of the present work. First, it may be of interest to compare our approach with further data-driven techniques, in particular with multivariate resampling techniques as proposed, for instance, by \cite{westfallyoung}. We have not included such a comparison here, because our main point was to demonstrate how much can be gained by choosing the blocks in a sophisticated manner instead of a naive choice. Second, one may consider nonparametric copula estimators in  $\hat{\mathbf{C}}_2$.  Finally, from the theoretical perspective it may of interest to analyze conditions for the validity of the MSM$_2$ property for certain relevant families of pair copulae. In the case of Archimedean copula families, an important contribution in this direction has been made by \cite{Scarsini2005}. The authors analyze conditions for the validity of the MTP$_2$ property for such families. It is well-known that MTP$_2$ distributions are also MSM$_2$ distributions; see \cite{glaz1984}.

\section*{Acknowledgments}
We thank Claudia Czado for fruitful discussions and Andr\'{e} Neumann for helpful comments.

\appendix


\section{Proof of Lemma \ref{lemma31}}\label{sec7}
Straightforward calculation yields, that
\begin{align}
\gamma_{j,i}(\boldsymbol{c})&=\mathbb{P}_{\boldsymbol\vartheta^*,C_\mathbf{X}} \left(T_j \leq c_j \mid \bigcap_{h=j-i+1}^{j-1} \{T_h \leq c_h \} \right)\nonumber\\
&=\frac{\mathbb{P}_{\boldsymbol\vartheta^*,C_\mathbf{X}} \left(\bigcap_{h=j-i+1}^{j} \{T_h \leq c_h \} \right)}{\mathbb{P}_{\boldsymbol\vartheta^*,C_\mathbf{X}} \left(\bigcap_{h=j-i+1}^{j-1} \{T_h \leq c_h \} \right)}\nonumber\\
&=\frac{\mathbb{P}_{\boldsymbol\vartheta^*,C_\mathbf{X}} \left(\bigcap_{h=j-i+1}^{j} \{F_{T_h}(T_h) \leq F_{T_h}(c_h) \} \right)}{\mathbb{P}_{\boldsymbol\vartheta^*,C_\mathbf{X}} \left(\bigcap_{h=j-i+1}^{j-1} \{F_{T_h}(T_h) \leq F_{T_h}(c_h) \} \right)}, \label{aha1}
\end{align}
because $F_{T_h}$ is assumed strictly increasing. Define $U_h =F_{T_h}(T_h)$. By the principle of probability integral transform, this random variable is uniformly distributed on $[0,1]$ under $\boldsymbol\vartheta^*$. Hence, we get that
\begin{eqnarray*}
\mathbb{P}_{\boldsymbol\vartheta^*,C_\mathbf{X}} \left(\bigcap_{h=j-i+1}^{j} \{F_{T_h}(T_h) \leq F_{T_h}(c_h) \} \right)&=& \mathbb{P}_{\boldsymbol\vartheta^*,C_\mathbf{X}} \left(\bigcap_{h=j-i+1}^{j} \{U_h \leq 1- \alpha_{loc} \} \right)\\
&=&C_{T_{j-i+1},...,T_{j}}(1- \alpha_{loc},...,1- \alpha_{loc}),
\end{eqnarray*}
by definition of $c_h$. Applying the analogous calculation to the denominator of \eqref{aha1} yields the assertion.

\bibliographystyle{elsarticle-harv}
\bibliography{vine-meff-arxiv}

\end{document}